\documentclass[sigconf]{acmart}

\usepackage{subfigure}

\usepackage{amssymb}
\usepackage{multirow}
\usepackage{graphicx}
\usepackage{float}
\usepackage{mathtools}
\usepackage{wrapfig}
\usepackage{setspace}
\usepackage{bbding}
\usepackage{pifont}
\usepackage{enumitem}

\acmConference[Woodstock '18]{Woodstock '18: ACM Symposium on Neural
  Gaze Detection}{June 03--05, 2018}{Woodstock, NY}
\acmBooktitle{Woodstock '18: ACM Symposium on Neural Gaze Detection,
  June 03--05, 2018, Woodstock, NY}
\acmPrice{15.00}
\acmISBN{978-1-4503-XXXX-X/18/06}

\begin{document}

\title{Geometric Interaction Augmented Graph Collaborative Filtering}

\author{Yiding Zhang}
\email{zy@bupt.edu.cn}
\affiliation{%
	\institution{Microsoft}
	\state{Beijing}
	\country{China}}
	
\author{Chaozhuo Li}\authornote{Equal contribution and corresponding author.}
\email{cli@microsoft.com}
\affiliation{%
	\institution{Microsoft Research Asia}
	\state{Beijing}
	\country{China}}
	
\author{Senzhang Wang} 
\email{szwang@csu.edu.cn} 
\affiliation{ 
	\institution{Central South University}
	\state{Changsha}
	\country{China}}

\author{Jianxun Lian, Xing Xie} 
\email{{jialia,xingx}@microsoft.com} 
\affiliation{ 
	\institution{Microsoft}
	\state{Beijing}
	\country{China}}


\begin{abstract}
Graph-based collaborative filtering is capable of capturing the essential and abundant collaborative signals from the high-order interactions, and thus received increasingly research interests. 
Conventionally, the embeddings of users and items are defined in the Euclidean spaces, along with the propagation on the interaction graphs. 
Meanwhile, recent works point out that the high-order interactions naturally form up the tree-likeness structures, which the hyperbolic models thrive on. 
However, the interaction graphs inherently exhibit the hybrid and nested geometric characteristics, while the existing single geometry-based models are inadequate to fully capture such sophisticated topological patterns. 
In this paper, we propose to model the user-item interactions in a hybrid geometric space, in which the merits of Euclidean and hyperbolic spaces are simultaneously enjoyed to learn expressive representations. 
Experimental results on public datasets validate the effectiveness of our proposal. 
\end{abstract}



\maketitle

\section{Introduction}
Personalized recommendation \cite{isinkaye2015recommendation} has been widely deployed to alleviate the challenge of information overload, which aims to predict whether a user will interact with an item. 
Collaborative filtering (CF) tries to address this problem on the basis of assuming behaviorally similar users would exhibit similar preference on items \cite{su2009survey}. 
As a classic recommendation paradigm, CF has been widely used in a myriad of scenarios such as E-commerce \cite{schafer2001commerce,sarwar2000analysis} and social recommendations \cite{tang2013social}. 

\begin{figure}
	\centering
	\subfigure[\scriptsize User-item relation ]{
		\includegraphics[width=0.18\textwidth]{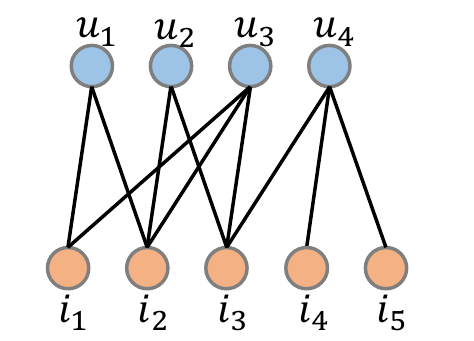}
		\label{fig_recommender_graph}
	}
	\subfigure[\scriptsize Tree-likeness interaction graph for $u_1$]{
		\includegraphics[width=0.24\textwidth]{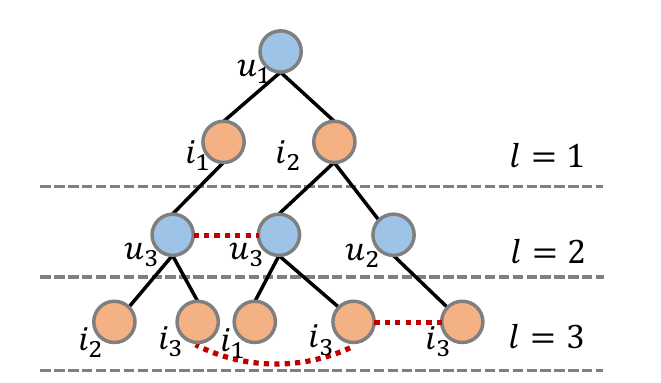}
		\label{fig_recommender_tree}
	}
	\vspace{-2mm}
	\caption{An illustration of the user-item interactions and the tree-likeness structures.}
	\vspace{-2mm}
	\label{fig_recommender}
\end{figure}

In order to exploit abundant collaborative signals, recent CF works \cite{wang2019neural,he2020lightgcn} propose to incorporate the high-order connectivity on the user-item interaction graphs, which contributes to facilitating the recommendation performance. 
As shown in Figure \ref{fig_recommender}, target user $u_{1}$ perceives high-order interactions through the $l$ stacked order-wise propagation layers. 
The high-order interactions of user $u_1$ can be naturally extended to a tree-likeness structure in Figure \ref{fig_recommender_tree} based on the interactions in Figure  \ref{fig_recommender_graph} .   
This is reasonable as the receptive field tends to be exponentially larger in the higher orders \cite{sankar2021graph,zhang2022geometric}.  

Conventional CF models usually formulate the learning of user and item representations in the Euclidean spaces \cite{koren2022advances,he2017neural,wang2022localized}. 
However, the Euclidean spaces may have a large distortion when learning graphs with tree-likeness structures as shown in Figure \ref{fig_recommender_tree}, while hyperbolic spaces can achieve better performance with much lower distortions \cite{ganea2018hyperbolic}. 
The reason lies in that hyperbolic spaces expand faster than Euclidean spaces.
Considering a disk in a 2-dimensional hyperbolic space with constant curvature $-1$, 
the perimeter and area of the disk of hyperbolic radius $r$ 
are given as $2\pi\sinh r$ and $2\pi(\cosh r-1)$ respectively, and both of them grow exponentially with $r$. 
Inspired by such merits, some researchers propose to design CF models in hyperbolic spaces \cite{vinh2020hyperml,feng2020hme}. HyperMF \cite{vinh2020hyperml} explores metric learning in hyperbolic spaces for recommendation. 
HME \cite{feng2020hme} studies the next-POI recommendation in hyperbolic spaces, which captures sequential transition, user preference, category and region information in a shared hyperbolic space.

%
%
%
%

Although hyperbolic-based models demonstrate promising performance, such single geometry-based approaches might be insufficient to fully capture the hybrid and nested geometrical structures in the interaction graphs. 
As shown in \ref{fig_recommender_tree}, there exist several circles  such as $u_1\rightarrow i_1 \rightarrow u_3 \rightarrow i_3 \rightarrow u_2\rightarrow i_2 \rightarrow u_1$. 
Such cyclic structures are prevalent in the interaction graph since they preserve the crucial collaborative signals. 
The cyclic and tree structures are nested together, leading to a hybrid geometric graph. 
Single geometry-based approaches only capture partial information and thus cannot depict user preferences from the holistic perspective.  

In this paper, we propose a novel geometric graph collaborative filtering (GGCF) algorithm,
which leverages both Euclidean spaces and hyperbolic spaces to encode the user-item interaction graphs on the basis of  graph convolutional networks (GCN).
The graph convolution layer of GGCF is capable of learning the user and item representations in both Euclidean and hyperbolic spaces simultaneously. 
The cross-geometry features are further integrated via an interaction learning module to adjust the conformal structural information.
GGCF predicts the user-item relations based on the results in both Euclidean and hyperbolic spaces. 
Therefore, our proposal organically enjoys the merits of Euclidean and hyperbolic spaces to advance the model performance.  

To summarize,
we make the following main contributions:
\begin{itemize}
	\item We propose to leverage Euclidean and hyperbolic spaces simultaneously in recommender systems, and design a novel geometry graph collaborative filtering method 
	to better capture the complex geometric patterns in  interaction graphs.
	\item Experimental results compared with state-of-the-art baselines verify the effectiveness of the proposed method.
\end{itemize}

\section{Preliminaries}
\subsection{Hyperbolic geometry}
Hyperbolic geometry is a non-Euclidean geometry with a constant negative curvature.
The hyperboloid model, as a typical equivalent model describing the hyperbolic geometry, 
has been widely investigated  \cite{nickel2018learning,chami2019hyperbolic,liu2019hyperbolic,law2019lorentzian}.
Since the basic operations in the hyperbolic spaces are different from those in the Euclidean spaces,
here we brief introduce some hyperbolic operations used in this paper.
Let  $\mathbf{x},\mathbf{y}\in \mathbb{R}^{n+1}$,
then the \textbf{Lorentzian scalar product} is defined as:
$
\langle \mathbf{x},\mathbf{y} \rangle_\mathcal{L} := -x_0y_0+\sum_{i=1}^{n}x_i y_i.
$
We denote $\mathbb{H}^{n}$ as the $n$-dimensional hyperboloid manifold with constant negative curvature $-1$:
$
\mathbb{H}^{n}:=\{\mathbf{x}\in \mathbb{R}^{n+1}: \langle \mathbf{x}, \mathbf{x} \rangle_\mathcal{L}=-1, x_0>0 \}.
$

The \textbf{tangent space} is a useful to perform Euclidean operations undefined in hyperbolic spaces. The tangent space
 at $\mathbf{x}$ is defined as a $n$-dimensional vector space approximating $\mathbb{H}^{n}$ around $\mathbf{x}$: 
\begin{equation}\label{eq:tangent_space}
	\mathcal{T}_\mathbf{x}\mathbb{H}^{n}:=\{\mathbf{v}\in\mathbb{R}^{n+1}:\langle \mathbf{v}, \mathbf{x} \rangle_\mathcal{L}=0\}.
\end{equation}
Besides, for $\mathbf{v,w}\in \mathcal{T}_\mathbf{x}\mathbb{H}^{n}$, a Riemannian metric tensor is given as $g_\mathbf{x}(\mathbf{v,w}):=\langle \mathbf{v,w} \rangle_\mathcal{L} $.
Then the hyperboloid model is defined as the hyperboloid manifold $\mathbb{H}^{n}$ equipped with the Riemannian metric tensor $g_\mathbf{x}$.
The mapping between hyperbolic spaces and tangent spaces can be implemented  by \textbf{exponential map}  and \textbf{logarithmic map}.
The exponential map is a map from subset of a tangent space of $\mathbb{H}^{n}$  (i.e., $\mathcal{T}_\mathbf{x}\mathbb{H}^{n}$) to $\mathbb{H}^{n}$ itself.
The logarithmic map is the reverse map that maps back to the tangent space.
For points $\mathbf{x,y}\in \mathbb{H}^{n}, \mathbf{v}\in \mathcal{T}_\mathbf{x}\mathbb{H}^{n}$, 
such that $\mathbf{v}\neq\mathbf{0}=(1,0,\cdots,0)$ and $\mathbf{x}\neq \mathbf{y}$, 
exponential map $\exp_\mathbf{x}(\cdot)$ and logarithmic map $\log_\mathbf{x}
(\cdot)$ are defined as follows:  
\begin{flalign}\label{eq:exp_map}
	\exp_\mathbf{x}(\mathbf{v})&=\cosh({\|\mathbf{v}\|_\mathcal{L}})\mathbf{x}+\sinh({\|\mathbf{v}\|_\mathcal{L}})\frac{\mathbf{v}}{\|\mathbf{v}\|_\mathcal{L}},\\
	\log_\mathbf{x}^\beta(\mathbf{y})&=
	d_\mathbb{H}(\mathbf{x,y})
	\frac{\mathbf{y}+\langle\mathbf{x,y}\rangle_\mathcal{L}\mathbf{x}}
	{\|\mathbf{y}+\langle\mathbf{x,y}\rangle_\mathcal{L}\mathbf{x}\|_\mathcal{L}},
\end{flalign}
where $\|\mathbf{v}\|_\mathcal{L}=\sqrt{\langle \mathbf{v},\mathbf{v} \rangle_\mathcal{L}}$ denotes Lorentzian norm of $\mathbf{v}$ and $d_\mathbb{H}(\cdot,\cdot)$ is the intrinsic distance function between two points $\mathbf{x},\mathbf{y} \in \mathbb{H}^{n}$:
\begin{equation}\label{eq:intrinsic_distance}
	d_\mathbb{H}(\textbf{x}, \textbf{y}) = \rm{\ arcosh}(-{\langle\mathbf{x},\mathbf{y} \rangle_\mathcal{L}} ).	
\end{equation}

\section{Methodology}
\begin{figure*}[htb]
	\centering
	\includegraphics[width=0.88\textwidth]{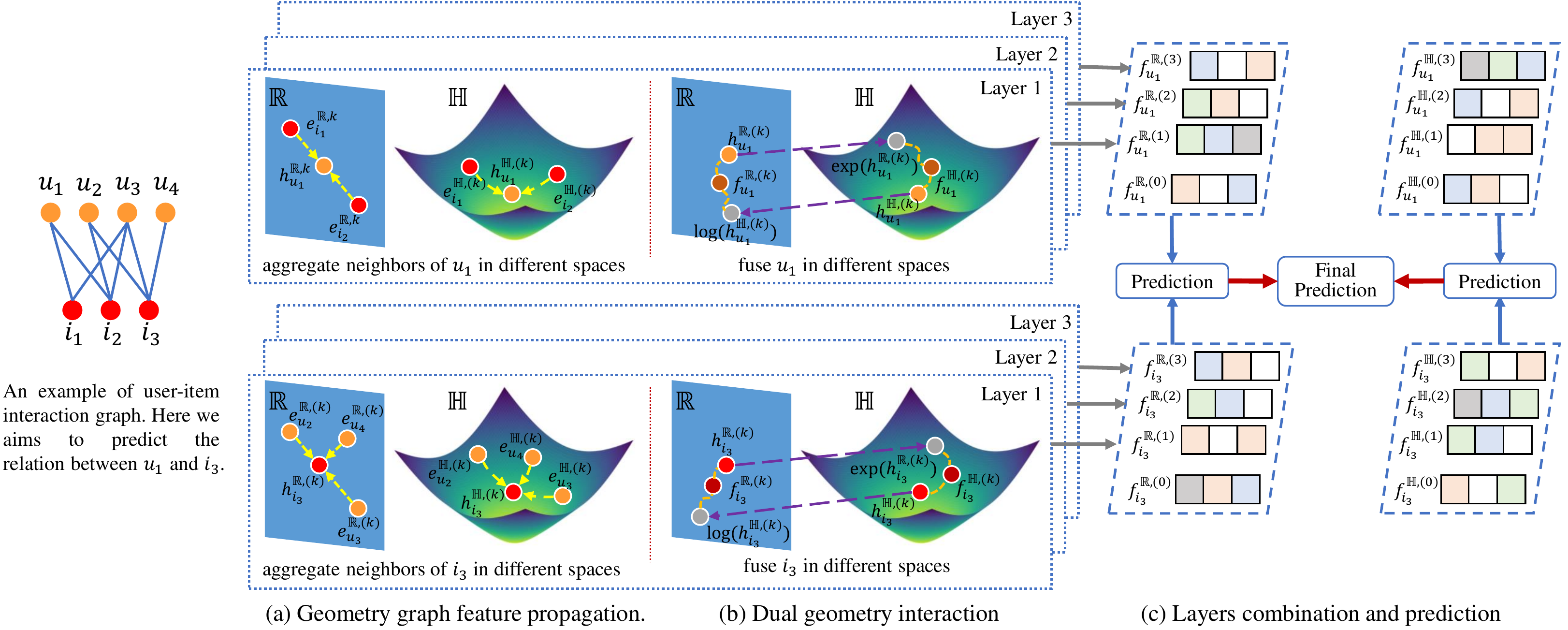}
	\caption{An illustration of the proposed model architecture.}
	\label{fig_model}
\end{figure*}

Figure \ref{fig_model} illustrates the framework of the  proposed GGCF model, which includes the following three modules:
\begin{itemize}[leftmargin=*]
	\item \textbf{Geometric feature propagation.}
	This module models the interactions between users and items by feeding these interactions into simplified graph convolutional layers, which can learn the graph representations in both hyperbolic and Euclidean spaces.
	
	\item \textbf{Dual geometry interaction.}
	This module fuses the features in hyperbolic and Euclidean spaces with each other, to combine different characteristic in different spaces.

	\item \textbf{Layers fusion and prediction.}
	Each layer has its own features of the user and items. Here we fuse the features in different layers to obtain more powerful representations, and predict the relations of users and items.
	
\end{itemize}
	

\subsection{Geometric feature propagation}

Graph convolutional networks have shown their powerful ability in recommender systems \cite{he2020lightgcn}.
Inspired by LightGCN \cite{he2020lightgcn} which observes feature transformation and nonlinear activation in GCNs could be burdensome for collaborative filtering,
we adopt the simple neighbor aggregator in our method.
For $k$-th graph convolutional layer, 
given input embedding $\mathbf{e}_u^{\kappa, (k)}$ and $\mathbf{e}_i^{\kappa, (k)}$ 
for user $u$ and item $i$ in Euclidean $\mathbb{R}$ or hyperbolic $\mathbb{H}$ spaces ($\kappa\in\{\mathbb{R},\mathbb{H}\}$),
we obtain embeddings by aggregating the features of neighbors:
$\mathbf{h}_u^{\kappa, (k)} = \text{Agg}^\kappa_{i\in\mathcal{N}_u}(w_{ui}, \mathbf{e}_i^{\kappa, (k)}),
		\mathbf{h}_i^{\kappa, (k)} = \text{Agg}^\kappa_{u\in\mathcal{N}_i}(w_{ui}, \mathbf{e}_u^{\kappa, (k)}),$
where 
$N_u$, $N_i$ are the set of neighbors of $u$ and $i$, respectively, 
and
$w_{ui}={1}/({\sqrt{|\mathcal{N}_u|}\sqrt{|\mathcal{N}_i|}})$ is denoted as the weight between $u$ and $i$.
The aggregator $\text{Agg}^{\kappa}$ in Euclidean space is defined as Euclidean weighted mean:
$
	\text{Agg}^\mathbb{R}_{i\in\mathcal{N}_u}(w_{ui}, \mathbf{e}_u^{\mathbb{R}, (k)}) = \sum_{i\in\mathcal{N}_u}w_{ui}\mathbf{e}_i^{\mathbb{R}, (k)}=\mathbf{h}_u^{\mathbb{R}, (k)}.
$
The weighted mean in non-Euclidean space is defined as  
Fr\'echet mean \cite{frechet1948elements,karcher1987riemannian,karcher2014riemannian},
which minimizes an expectation of (squared) distances with a set of points.
However,
Fr\'echet mean does not have a closed form solution,
and it cannot be efficiently calculated by stochastic gradient descent. 
Thus, we leverage an elegant neighborhood aggregation method based on the centroid \cite{law2019lorentzian}:
\begin{equation}\label{eq:agg_hyperboloid}
	\text{Agg}^\mathbb{H}_{i\in\mathcal{N}_u}(w_{ui}, \mathbf{e}_i^{\mathbb{H}, (k)}) 
	=\frac{\sum_{i\in\mathcal{N}_u}w_{ui}\mathbf{e}_i^{\mathbb{H}, (k)}}{|\|\sum_{i\in\mathcal{N}_u}w_{ui}\mathbf{e}_i^{\mathbb{H}, (k)}\|_{\mathcal{L}}|}=\mathbf{h}_i^{\mathbb{H}, (k)}.
\end{equation}
Thus, the features $\mathbf{h}_i^{\mathbb{R}, (k)}$ and  $\mathbf{h}_i^{\mathbb{H}, (k)}$ can be easily obtained by  aggregating their neighbors.

\subsection{Dual geometry interaction}
The geometric feature propagation is built upon the graph topology.
Here we aim to verify that the user/item features are consistent with the dual feature spaces to promote interaction between Euclidean and hyperbolic geometries.
A straight-forward strategy is to transform the features between hyperbolic and Euclidean spaces and compute the average features directly.
However, this strategy cannot preserve the difference between features in these two types of spaces, and thus we should fuse the features in an adaptive manner. 
Therefore, we propose to adjust the features in different spaces based on their similarities.
Specifically, given Euclidean feature $\mathbf{h}_i^{\mathbb{R}, (k)}$ and hyperbolic feature $\mathbf{h}_i^{\mathbb{H}, (k)}$ of item $i$,
we leverage their distance to evaluate their similarities and fuse the features as:
\begin{equation}
\begin{split}
	\mathbf{f}_i^{\mathbb{R},(k)}&=\mathbf{h}_i^{\mathbb{R},(k)}+
	\big(\gamma d_\mathbb{R}(\mathbf{h}_i^{\mathbb{R},(k)},\log_\mathbf{0}(\mathbf{h}_i^{\mathbb{H},(k)}))\times\log_\mathbf{0}(\mathbf{h}_i^{\mathbb{H},(k)})\big),\\
	\mathbf{f}_i^{\mathbb{H},(k)}&=\mathbf{h}_i^{\mathbb{H},(k)}\oplus
	\big(\gamma' d_\mathbb{H}(\mathbf{h}_i^{\mathbb{H},(k)},\exp_\mathbf{0}(\mathbf{h}_i^{\mathbb{R},(k)}))\otimes\exp_\mathbf{0}(\mathbf{h}_i^{\mathbb{R},(k)})\big),
\end{split}
\end{equation}
where $\exp_\mathbf{0}(\cdot)$ and $\log_\mathbf{0}(\cdot)$ are leveraged to transform the features between Euclidean and hyperbolic spaces.
The similarity from different spatial features is measured by Euclidean distance $d_\mathbb{R}$ hyperbolic distance $d_\mathbb{H}$ in Eq. \eqref{eq:intrinsic_distance}, scaled by trainable scalar $\beta, \beta'$. 
$\oplus$ and $\otimes$ are addiction and scalar multiplication defined in hyperbolic spaces.
Given feature vectors $\mathbf{x}, \mathbf{y}\in\mathbb{H}, \mathbf{v}\in\mathcal{T}_\mathbf{x}\mathbb{H}$, and a scalar $r$, these multiplications are defined as:
$\mathbf{x}\oplus\mathbf{y}=\exp_\mathbf{x}(P_{\mathbf{0}\rightarrow\mathbf{x}}(\log_\mathbf{0}(\mathbf{y}))), P_{\mathbf{0}\rightarrow\mathbf{x}}(\mathbf{v})=\mathbf{v}+(\mathbf{0}+\mathbf{x})\cdot\langle\mathbf{x},\mathbf{v}\rangle_\mathcal{L}/(1-\langle\mathbf{0},\mathbf{x}\rangle_\mathcal{L}),$
and
$r\otimes\mathbf{x}=\exp_\mathbf{0}(r\cdot\log_\mathbf{0}(\mathbf{x})).$

\subsection{Layers fusion and prediction}
After applying $K$ geometric graph convolutional layers,
we can obtain features in each layer.
To obtain more powerful representations, we fuse all the features in each layer along with the input features (e.g., the $0$-th layer) through the equal weight $1/(K+1)$:
\begin{equation}
	\mathbf{f}_i^{\mathbb{R}}=\sum_{k=0}^K {\mathbf{f}_i^{\mathbb{R},(k)}}/(K+1),\quad
	\mathbf{f}_i^{\mathbb{H}}=\sum_{k=0}^K
	\frac{{\mathbf{f}_i^{\mathbb{H},(k)}}/(K+1)}{|\|{\mathbf{f}_i^{\mathbb{H},(k)}}/(K+1)\|_\mathcal{L}|}. 
\end{equation}
Thus, we can prediction the relation between user and item with final feature representations with Euclidean and Lorentzian scalar product:
$
	\hat{y}_{ui}=\langle \mathbf{f}_u^{\mathbb{R}},\mathbf{f}_i^{\mathbb{R}} \rangle + \lambda \langle \mathbf{f}_u^{\mathbb{H}},\mathbf{f}_i^{\mathbb{H}} \rangle_\mathcal{L},
$
where $\lambda$ is a trainable parameter to balance the importance between the Euclidean and hyperbolic parts in our model.

\subsection{Model training}
We employ the \textit{Bayesian Personalized Ranking} (BPR) loss \cite{rendle2012bpr},
which encourages the prediction of an observed entry to be higher than its unobserved counterparts:
$
	\mathcal{L} = - \sum_{u}\sum_{i\in\mathcal{N}_u}\sum_{j\notin\mathcal{N}_u}\ln \sigma(\hat{y}_{ui}-\hat{y}_{uj}).
$
We leverage $L_2$ regularization to avoid overfitting.
Please note that the input features for our hyperbolic graph convolutional layers are initialized in the Euclidean space $\mathcal{T}_\mathbf{0}\mathbb{H}$ and mapped to hyperbolic spaces via $\exp_\mathbf{0}(\cdot)$.
In the process of transforming features between hyperbolic and Euclidean (tangent) spaces,
we concatenate $``0"$ and the features to satisfy the definition in Eq. \eqref{eq:tangent_space}.
Since all the trainable parameters are defined in Euclidean spaces, we employ the Adam \cite{kingma2014adam} optimizer and train our model in a mini-batch manner.

\subsection{Discussion}
\subsubsection{Relation with GIL}
The graph convolutional layer in GIL \cite{zhu2020graph} is  different from our work. 
GIL employs feature transformation and non-linear activation,
while our proposal discards such operations since they might be burdensome for collaborative filtering \cite{he2020lightgcn}.
Meanwhile, the hyperbolic neighbor aggregator in GIL is different from ours. 
The aggregator in our method is based on the Lorentzian centorid, while the one in GIL is based on the tangent spaces.

\section{Experiments}
In this section, we conduct extensive experiments on publicly available datasets to answer the following three research questions:
\begin{itemize}[leftmargin=*]
	\item \textbf{RQ1}:
	How dose the proposed GGCF model perform as compared with the state-of-the-art methods?
	
	\item \textbf{RQ2}:
	What are the key components that influence the model performance? 
	
	\item \textbf{RQ3}:
	How do different hyper-parameter settings (e.g., the number of graph convolutional layers) affect the  model performance?
\end{itemize}

\subsection{Experimental Settings}
\subsubsection{Datasets and Metrics}
We select two popular real-world datasets for evaluation. 
Table \ref{table_dataset} presents the statistical details. 
{Movielens}\footnote{https://grouplens.org/datasets/movielens/}
describes the ratings and free-text tagging activities on the MovieLens website. 
{LastFM}\footnote{http://www.lastfm.com} contains the interactions between the users and artists. 
We follow the settings in NGCF \cite{wang2019neural} to split the training/test sets for a fair comparison. 
Following the previous works \cite{wang2019neural,he2020lightgcn}, we leverage recall@20 and ndcg@20 to evaluate the effectiveness of our proposal.

\begin{table}[t]
	\caption{Statistics of the  datasets.}
	\label{table_dataset}
	\centering
	\begin{tabular}{cccc}
		\toprule
		Dataset		& \# User	&	\# Item &	\# Interaction 	\\
		\midrule
		MovieLens	&	610		&	9,742	&	100,836 	\\
		LastFM		&	1,892 	&	17,632	&	92,834 	\\
		\bottomrule
	\end{tabular}
\end{table}

\subsubsection{Baselines} 
We select the following SOTA CF models as baselines to demonstrate the superiority of GGCF model: 
\begin{itemize}[leftmargin=*]
	\item BPRMF \cite{rendle2012bpr} is a matrix factorization based CF model under the framework of Bayesian personalized ranking.
	
	\item HyperML \cite{vinh2020hyperml} is a hyperbolic CF model, which explores metric learning in hyperbolic spaces.
	
	\item NGCF \cite{wang2019neural} is a graph-based CF model to incorporate the high-order connectivity of user-item interactions.
	
	\item LightGCN \cite{he2020lightgcn} is a state-of-the-art CF recommendation model based on light graph convolution.
\end{itemize}

The dimension of user/item embeddings is set to 64. 
The number of graph convolutional layers in graph-based models is set to 3. 
Parameters are carefully tuned on a small validation dataset. 
Specifically, the learning rate is searched in $\{1e-2, 5e-3, 1e-3, 5e-4, 1e-4\}$ and $L_2$ regularization term is tuned in $\{0, 1e-6, 1e-5, 1e-4, 1e-3\}$.
%

\subsection{Overall Performance (RQ1)}

\begin{table}[t]
	\caption{Overall Performance.}
	\label{table_preformance}
	\centering
	\begin{tabular}{ccccc}
		\toprule
		Dataset		& \multicolumn{2}{c}{MovieLens}	&	\multicolumn{2}{c}{LastFM} \\
				& recall@20	&	ndcg@20		&	recall@20	&	ndcg@20	\\
		\midrule
		BPRMF		& 0.1914	&	0.2561		&	0.2180		&	0.2106		\\
		hyperML		& 0.2076	&	0.2654		&	0.2164		&	0.2094		\\
		NGCF		& 0.2197	&	0.3164		&	0.2374		&	0.2409		\\
		LightGCN	& 0.2517	&	0.3362		&	0.2592		&	0.2610		\\
		GGCF		& \textbf{0.2589}&\textbf{0.3420}&\textbf{0.2621}		&\textbf{0.2691}		\\
		\bottomrule
	\end{tabular}
\end{table}

The experiment results are presented in Table \ref{table_preformance}. 
%
From the results, we can obtain the following observations.
\begin{itemize}[leftmargin=*]
	\item \textbf{Our proposed GGCF model achieves the best performance.} 
	GGCF consistently outperforms all the baseline methods on both datasets under different evaluation metrics, which  indicates the superiority of simultaneously modeling user-item relations in the Euclidean and hyperbolic spaces.

	\item \textbf{Graph-based models are more powerful.}
	The graph-based models (i.e., NGCF, LightGCN and GGCF) outperform the traditional shallow models (i.e., BPRMF and hyperML),
	which verifies the benefit of incorporating high-order topological connectivity as complementary.

	\item \textbf{Single hyperbolic space is not enough.} 
	One can see that the pure hyperbolic CF model (hyperML) performs worse than the Euclidean-based approaches (e.g, NGCF and LightGCN), which suggests that the single  hyperbolic space may be insufficient to capture the sophisticated graph topology. However, by enjoying the merits from both Euclidean and hyperbolic sides, the performance of our proposal surpasses the single space-based models. 
	It demonstrates that GGCF is capable of capturing the cross-space correlations to learn more comprehensive user/item representations.      
	
\end{itemize}

\subsection{Ablation Study (RQ2)}

\begin{table}[t]
	\caption{Experimental results of the ablation Study.}
	\label{table_ablation}
	\centering
	\begin{tabular}{ccccc}
		\toprule
		Dataset		& \multicolumn{2}{c}{MovieLens}	&	\multicolumn{2}{c}{LastFM} \\
				& recall@20	&	ndcg@20		&	recall@20	&	ndcg@20	\\
		\midrule
		GGCF$_{\backslash\text{I}}$& 0.2509	&	0.3295	&	0.2602	&	0.2630		\\
		GGCF$_{\backslash\text{E}}$& 0.2412	&	0.3160	& \textbf{0.2641}	&	0.2517	\\
		GGCF$_{\backslash\text{H}}$& 0.2514	&	0.3359	& {0.2598}	&	0.2623	\\
		GGCF		& \textbf{0.2589}&\textbf{0.3420}&{0.2621}		&\textbf{0.2691}		\\
		\bottomrule
	\end{tabular}
\end{table}

In order to evaluate the effectiveness of critical components in GGCF, we design the following three ablation models: GGCF$_{\backslash\text{E}}$ without the hyperbolic part, GGCF$_{\backslash\text{H}}$ without the Euclidean part and the ensemble model GGCF$_{\backslash\text{E}}$ without the interaction components.  
Experimental results are shown in Table \ref{table_ablation}.

\begin{itemize}[leftmargin=*]
	\item \textbf{Effectiveness of the cross-space interactions.}
	GGCF outperforms the ablation model GGCF$_{\backslash\text{I}}$ by nearly 1.3\% on average, which indicates that the interactions between the Euclidean and hyperbolic spaces can further enrich the complementary relationships.
	
	\item \textbf{Effectiveness of Euclidean and hyperbolic graph convolutional part.}
	Compared with GGCF$_{\backslash\text{E}}$ and GGCF$_{\backslash\text{H}}$ models,
	GGCF performs better in most cases, which also indicates GGCF can make a desirable balance between the information from Euclidean and hyperbolic spaces.
\end{itemize}

\subsection{Parameter Sensitivity Study (RQ3)}

\begin{figure}
	\centering
	\subfigure{
		\includegraphics[width=0.22\textwidth]{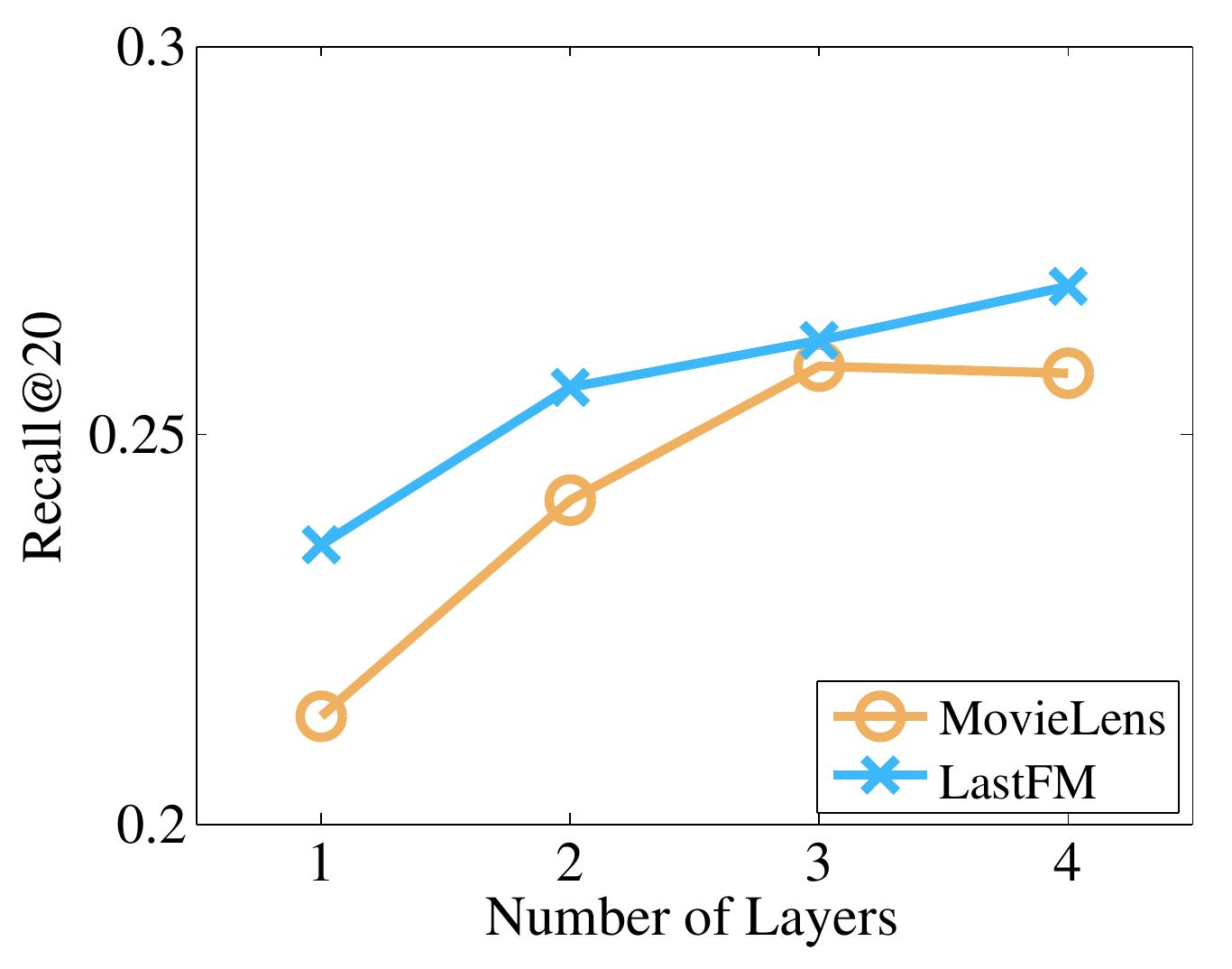}
		\label{fig_dim_recall}
	}
	\subfigure{
		\includegraphics[width=0.22\textwidth]{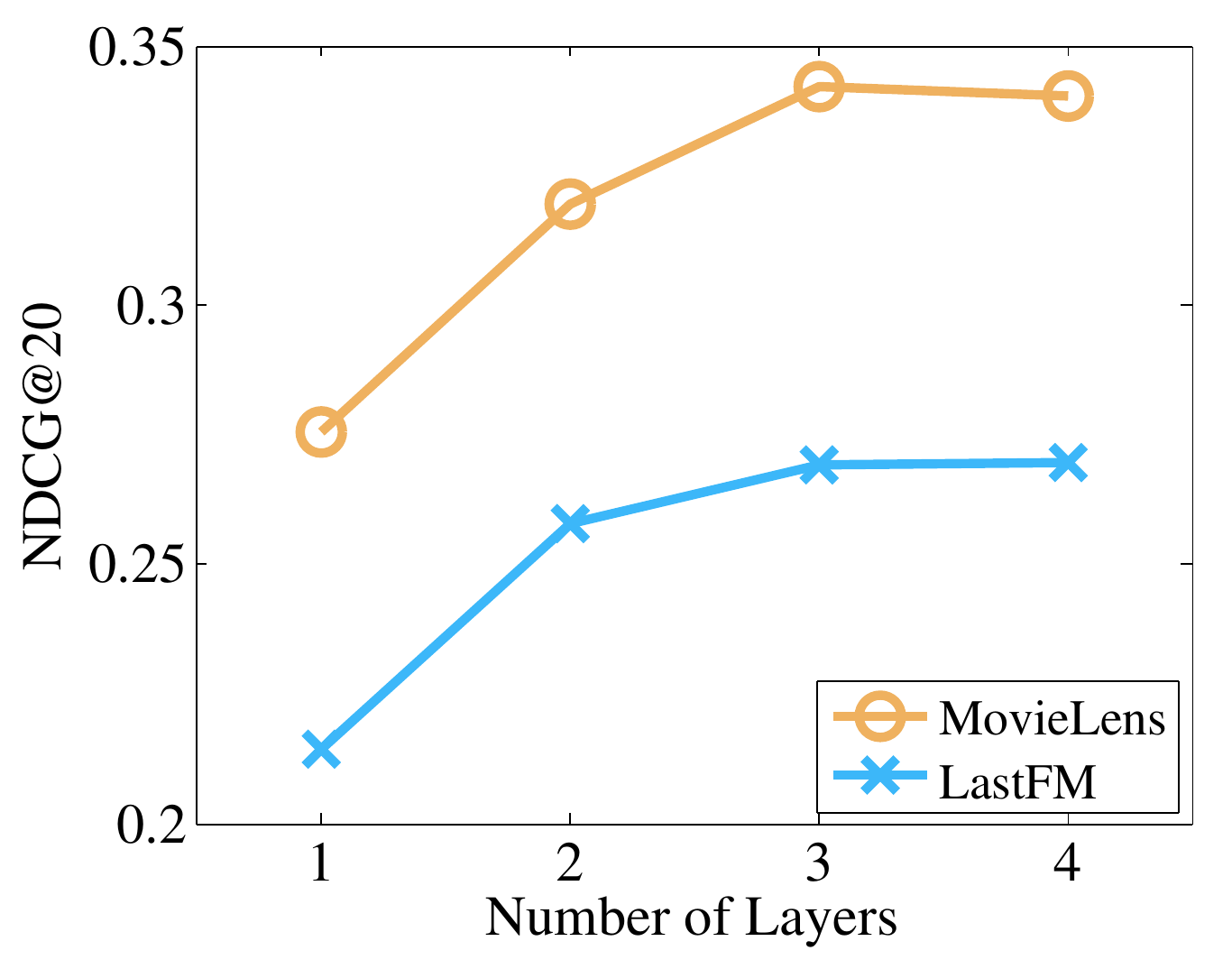}
		\label{fig_dim_ndcg}
	}
	\caption{Results of GGCF with different number of layers.}
	\label{fig_dim}
\end{figure}

Here we study the performance sensitivity of HAUIL model on the core parameter: the number of graph convolutional layer. 
We change the number of layers from 1 to 4, and the results are shown in Figure \ref{fig_dim}. 

With the increase of number of layers, the performance over almost all the datasets first increases and then keeps steady or slightly drops, which demonstrates appropriate high-order topological information may benefit the model performance. 
However,  a too large number of layers will lead to challenge of ``neighborhood explosion'', which may introduces noises and will be resource-consuming.  
Thus, we have to carefully choose an appropriate value to balance the model efficiency and effectiveness.  

\section{Conclusion}
In this work,
we study the novel problem of graph-based CF. 
We propose a novel geometry graph collaborative filtering method for recommender systems,
to learn the user/item representations in both Euclidean and hyperbolic spaces.
These representations are further enhanced by the dual geometry interaction scheme to take advantage of both Euclidean and hyperbolic spaces.
The experimental results on public datasets demonstrate the effectiveness of the propose method.

\bibliographystyle{ACM-Reference-Format}
\bibliography{sample-base}
\end{document}